\documentclass[letterpaper, 10 pt, conference]{ieeeconf}  
\IEEEoverridecommandlockouts                              
\usepackage{tabularx,xtab,array,float}
\usepackage{epsfig,epstopdf,wrapfig,psfrag}
\usepackage{amsmath,amssymb,graphicx,graphics}
\usepackage{tikz}
\usepackage{balance}
\usepackage{cite}
\usepackage{soul}
\usepackage{hyperref}

\makeatletter
\newcommand*{\encircled}[1]{\relax\ifmmode\mathpalette\@encircled@math{#1}\else\@encircled{#1}\fi}
\newcommand*{\@encircled@math}[2]{\@encircled{$\m@th#1#2$}}
\newcommand*{\@encircled}[1]{%
  \tikz[baseline,anchor=base]{\node[draw,circle,outer sep=0pt,inner sep=.2ex] {#1};}}
\makeatother

\newtheorem{remark}{Remark}

\newtheorem{assumption}{Assumption}

\begin{document}
%
\title{\LARGE \bf Stabilization of Energy-Conserving Gaits for Point-Foot Planar Bipeds}
%
%
%
\author{Aakash~Khandelwal, Nilay~Kant, and~Ranjan~Mukherjee
\thanks{Department of Mechanical Engineering, Michigan State University, East Lansing, MI 48824, USA}
}
	
\maketitle
\thispagestyle{empty}
\pagestyle{empty}

\begin{abstract}
The problem of designing and stabilizing impact-free, energy-conserving gaits is considered for underactuated, point-foot planar bipeds. Virtual holonomic constraints are used to design energy-conserving gaits. A desired gait corresponds to a periodic hybrid orbit and is stabilized using the Impulse Controlled Poincar\'e Map approach. Numerical simulations for the case of a five-link biped demonstrate convergence to a desired gait from arbitrary initial conditions.
\end{abstract}

\section{Introduction} \label{sec1}

Point-foot bipeds represent a class of underactuated systems whose dynamics are hybrid due to foot-ground interaction and interchange of stance and swing legs. An important control problem is to design and stabilize a biped gait, which is a periodic hybrid orbit. Virtual Holonomic Constraints (VHCs) \cite{grizzle_asymptotically_2001, shiriaev_constructive_2005, maggiore_virtual_2013} have been used to design gaits for bipeds; systematic selection of VHCs and parameter optimization have been used to guarantee stability of gaits \cite{grizzle_asymptotically_2001, westervelt_hybrid_2003, freidovich_stability_2008, chevallereau_asymptotically_2009, grizzle_virtual_2019, westervelt_feedback_2018, boroujeni_unified_2021}.\

In this work, we consider point-foot planar bipeds having $n$ degrees-of-freedom (DOFs) and $(n\!-\!1)$ control inputs. For this class of systems, we design and stabilize \emph{energy-conserving} gaits, that conserve mechanical energy of the biped over each step and are free from impact during foot-ground interaction. The notion of an energy-conserving gait appeared in \cite{jafari_energy-conserving_2013}; here it is formally described using VHCs, and designed to satisfy conditions for feasibility and stabilizability. Such gaits can be expressed as hybrid orbits that are invariant to coordinate relabelling. The Impulse Controlled Poincar\'e Map (ICPM) approach, which has been used to stabilize both continuous and hybrid orbits \cite{kant_orbital_2020, kant_juggling_2022}, is used for stabilizing a desired energy-conserving gait from arbitrary initial conditions. The efficacy of the control design is demonstrated through a case study of the five-link biped.\

\section{System Dynamics} \label{sec2}

\subsection{System Description and Assumptions} \label{sec21}

\begin{figure}[t!]
 \centering
 \psfrag{a}[][]{\small {$x$}}
 \psfrag{b}[][]{\small {$y$}}
 \psfrag{1}[][]{\footnotesize {$\theta_1$}}
 \psfrag{2}[][]{\footnotesize {$\theta_2$}}
 \psfrag{3}[][]{\footnotesize {$\theta_j$}\,\,\,\, \scriptsize{$j\!=\!(n\!+\!1)/2$}}
 \psfrag{5}[][]{\footnotesize {$\theta_n$}}
 \psfrag{A}[][]{\footnotesize {$\ell_1$}}
 \psfrag{B}[][]{\footnotesize {$\ell_2$}}
 \psfrag{C}[][]{\footnotesize {$\ell_j$}}
 \psfrag{D}[][]{\footnotesize {$\ell_n$}}
 \includegraphics[scale=0.33]{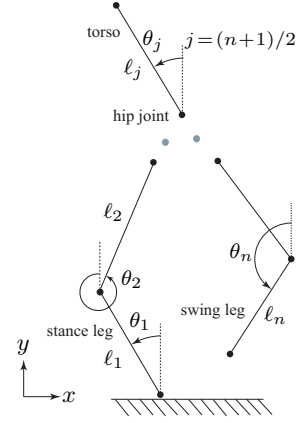}
 \caption{An $n$-link point-foot planar biped.}
 \label{Fig1}
 \vspace{-0.20in}
 \end{figure}

Consider the $n$-link planar biped ($n$ is odd) comprised of a single-link torso and kinematically similar legs with point feet - see Fig.\ref{Fig1}. It is assumed that the biped gait is comprised of a sequence of steps, where each step is comprised of a single-support phase (one leg is in contact with the ground) and a double-support phase (both legs are in contact with the ground). In the single-support phase, the leg in contact with the ground is referred to as the stance leg and the other leg is referred to as the swing leg. The stance foot is passive; it does not slide or leave the ground and acts as a frictionless pivot. The single-support phase ends when the swing leg comes in contact with the ground. The ensuing double-support phase is of infinitesimal time duration; during this phase, there is physical interaction between the ground and the swing leg but not between the ground and the stance leg. The double-support phase ends with relabelling of coordinates for interchange of the stance and swing legs.\

Each leg has $(n\!-\!1)/2$ links; the stance leg links are numbered sequentially $1$ through $(n\!-\!1)/2$ starting from the link in contact with the ground, the torso is link $(n\!+\!1)/2$, and the swing leg links are numbered sequentially $(n\!+\!3)/2$ through $n$ starting from the link in contact with the torso. The length of the $j$-th link, $j = 1, 2, \cdots, n$, is $\ell_j$. Since the legs are kinematically identical, the link lengths satisfy
\begin{equation*}
\ell_{n - j + 1} = \ell_j \quad \forall j = 1, 2, \cdots, (n\!-\!1)/2
\end{equation*}

\noindent The orientation of the $j$-th link, $j = 1, 2, \cdots, n$, measured counter-clockwise with respect to the vertical, is denoted by $\theta_j$, $\theta_j \in S^1$. The link $j$, $j = 2, \cdots, n$, is driven by an actuator mounted on link $(j\!-\!1)$, which applies torque $\tau_j$.\

The dynamics of the biped gait in the single-support phase, also known as the swing phase, is discussed in section \ref{sec22}. The dynamics of foot-ground interaction and the relabelling of coordinates for interchange of the stance and swing legs in the double-support phase is discussed in section \ref{sec23}. For generality and ease of control design, the dynamics are presented using the generalized coordinates $q \in \mathcal{Q}^n \triangleq \prod_{j = 1}^n S^1$, $q \triangleq \begin{bmatrix} q_1^T & q_2 \end{bmatrix}^T$, where $q_1 \in \mathcal{Q}^{n-1}$ and $q_2 \in S^1$. We define the generalized coordinates $q$ as follows
\begin{equation*}
q_1 = \begin{bmatrix} (\theta_2\!-\!\theta_1) &(\theta_3\!-\!\theta_2) &\cdots &(\theta_n\!-\!\theta_{n-1})\end{bmatrix}, \quad q_2 = \theta_1
\end{equation*}

\subsection{Swing Phase Dynamics}  \label{sec22}

During the swing phase, the biped represents an $n$-DOF underactuated system with one passive DOF $q_2$. The kinetic and potential energies of the system are denoted by $T(q, \dot q) = \frac{1}{2} \dot q^T M(q) \dot q$ and $V(q)$ respectively, where $M \in R^{n \times n}$ is the symmetric, positive definite mass matrix:
\begin{equation*}
M(q) = \begin{bmatrix} \begin{array}{c|c} 
M_{11}(q) & M_{12}(q) \\
\hline \\[-2.25ex]
M_{12}^T(q) & M_{22}(q)
\end{array} \end{bmatrix}
\end{equation*}

\noindent where $M_{11} \in R^{(n-1) \times (n-1)}$, $M_{22} \in R$ 
and the equations of motion can be written in the same form as in \cite{kant_orbital_2020}:
\begin{subequations} \label{eq1}
\begin{align}
M_{11}(q) \ddot{q}_1 + M_{12}(q) \ddot{q}_2 + h_1(q, \dot{q}) &= u \label{eq1a} \\
M_{12}^T(q) \ddot{q}_1 + M_{22}(q) \ddot{q}_2 + h_2(q, \dot{q}) &= 0 \label{eq1b}
\end{align}
\end{subequations}

\noindent where $\begin{bmatrix} h_1^T & h_2 \end{bmatrix}^T \in R^n$ is the vector of Coriolis, centrifugal, and gravity forces, and $u = \begin{bmatrix}\tau_2 &\tau_3 &\cdots &\tau_n \end{bmatrix}^T \in R^{n-1}$ is the vector of control inputs. Since the biped has revolute joints, we make the following assumption \cite{maggiore_virtual_2013, kant_orbital_2020}.

\begin{assumption} \label{asm1}
For the $n$-link biped, the mass matrix and the potential energy are even functions of $q$:
\begin{equation*}
M(q) = M(-q), \quad V(q)= V(-q)
\end{equation*}
which satisfies \cite[Assumption 1]{kant_orbital_2020}.
\end{assumption}

When $u$ is continuous and has the form $u = u_c(q, \dot q)$, the dynamics in \eqref{eq1} has the state-space representation
\begin{equation} \label{eq2}
\dot x = f(x), \qquad x \triangleq \begin{bmatrix} q^T & \dot q^T \end{bmatrix}^T \in \mathcal{Q}^n \times R^n
\end{equation}

\noindent If an impulsive input $u_\mathcal{I}$ is applied in addition to $u_c$ at any instant, the system will experience a discontinuous change in the generalized velocities with no change in the generalized coordinates \cite{kant_estimation_2019, kant_orbital_2020}. The jump in generalized velocities can be obtained by integrating \eqref{eq1} as follows:
\begin{equation} \label{eq3}
\begin{bmatrix}
M_{11} & M_{12} \\
M_{12}^T & M_{22}
\end{bmatrix}
\begin{bmatrix}
\Delta \dot q_1 \\ \Delta \dot q_2
\end{bmatrix} = 
\begin{bmatrix}
\mathcal{I} \\ 0
\end{bmatrix}, \quad \mathcal{I} \triangleq \int_0^{\Delta t} u_\mathcal{I} dt
\end{equation}

\noindent where $\Delta t$ is the infinitesimal duration for which $u_\mathcal{I}$ is active, $\mathcal{I} \in R^{n-1}$ is the impulse of $u_\mathcal{I}$,
\begin{equation*}
\Delta \dot q_1 \triangleq (\dot q_1^+ - \dot q_1^-), \quad \Delta \dot q_2 \triangleq (\dot q_2^+ - \dot q_2^-)
\end{equation*}

\noindent and $(.)^-$ and $(.)^+$ denote the variable $(.)$ immediately before and after an event where there is a discontinuous jump in its value. The states immediately after application of the impulsive input can be expressed as
\begin{equation}\label{eq4}
x^+ = x^- + \Delta x_{\mathcal{I}}, \quad \Delta x_{\mathcal{I}} \triangleq  \begin{bmatrix} 0 \\ \Delta \dot q \end{bmatrix}
\end{equation}

\noindent where $\Delta \dot q$ is obtained from \eqref{eq3}.\

\subsection{Foot-Ground Interaction and Coordinate Relabelling} \label{sec23}

In the double-support phase, there is impulsive interaction between the swing foot and the ground. Simultaneously, the stance foot lifts from the ground without interaction, and there is an instantaneous interchange between stance and swing legs.\

Following the approach in \cite{hurmuzlu_rigid_1994, grizzle_asymptotically_2001}, the impact between the swing foot and the ground is modeled as an inelastic collision. This model is described using $(n+2)$ DOF, that include the original $n$ DOFs and the two Cartesian coordinates of the stance foot $(s_x, s_y)$, $s_x, s_y \in R$. The equations of motion of the extended system can be written in the form
\begin{equation} \label{eq5}
M_e(q_e) \ddot{q}_e + h_e(q_e, \dot{q}_e) = p_e + p^{\rm ext}, \quad q_e \triangleq \begin{bmatrix} q^T &\!\!s_x &\!\!s_y \end{bmatrix}^T
\end{equation}

\noindent where $M_e \in R^{(n+2) \times (n+2)}$ is the mass matrix, $h_e \in R^{n+2}$ is the vector of Coriolis, centrifugal, and gravity forces, $p_e = \begin{bmatrix} u^T &0 &0 &0 \end{bmatrix}^T \in R^{n+2}$ is the vector of generalized forces, and $p^{\rm ext} \in R^{n+2}$ is the vector of generalized impulsive forces due to interaction between the swing foot and the ground. The discontinuous change in the generalized velocities due to $p^{\rm ext}$ can be obtained by integrating \eqref{eq5}
\begin{equation} \label{eq6}
M_e(q_e)\, \Delta \dot{q}_e  = \mathcal{I}^{\rm ext}, \quad \mathcal{I}^{\rm ext} \triangleq \int_{0}^{\Delta t} p^{\rm ext} dt
\end{equation}

\noindent where $\Delta t$ is the infinitesimal duration of the impact, $\mathcal{I}^{\rm ext} \in R^{n+2}$ is the impulse due to $p^{\rm ext}$, and $\Delta \dot{q}_e \triangleq (\dot{q}_e^+ - \dot{q}_e^-)$. Let $F_x$ and $F_y$ denote the Cartesian components of the impulsive forces on the swing foot due to impact. Then 
\begin{equation} \label{eq7}
p^{\rm ext} = \Gamma^T F, \quad \Gamma \triangleq \frac{\partial \gamma}{\partial q_e}, \quad \gamma \triangleq \begin{bmatrix} \gamma_x \\ \gamma_y \end{bmatrix}, \quad F \triangleq \begin{bmatrix} F_x \\ F_y \end{bmatrix}
\end{equation}

\noindent where $\gamma \equiv \gamma(q_e) \in R^2$ denotes the Cartesian coordinates of the swing foot. By integrating \eqref{eq7} over the duration of impact $\Delta t$, we get
\begin{equation} \label{eq8}
\mathcal{I}^{\rm ext} = \Gamma^T \mathcal{I}_{\rm g}, \quad \mathcal{I}_{\rm g} \triangleq \int_{0}^{\Delta t} F dt
\end{equation}

\noindent Since the foot-ground collision is inelastic,
\begin{equation} \label{eq9}
\Gamma \dot q_e^+ = 0
\end{equation}

\noindent Using \eqref{eq6}, \eqref{eq8} and \eqref{eq9}, we get
\begin{equation} \label{eq10}
\begin{bmatrix}
M_e & -\Gamma^T \\
\Gamma & 0
\end{bmatrix}
\begin{bmatrix}
\dot q_e^+ \\ \mathcal{I}_{\rm g}
\end{bmatrix} = \begin{bmatrix}
M_e \dot q_e^- \\ 0
\end{bmatrix}
\end{equation}

\noindent The states immediately after foot-ground interaction are
\begin{equation}\label{eq11}
x^+ = x^- + \Delta x_{\rm g}, \quad \Delta x_{\rm g} \triangleq  \begin{bmatrix} 0 \\ \Delta \dot q
\end{bmatrix}
\end{equation}

\noindent where $\Delta \dot q$ is obtained using \eqref{eq10}.\

The subsequent interchange of the stance and swing legs is equivalent to a relabelling of states \cite{grizzle_virtual_2019}. The states immediately after leg interchange are a function of those immediately before interchange, and is given by the relation:
\begin{align} 
x^{+} &= \mathcal{R}(x^-) \label{eq12} \\
\mathcal{R}(x) &= {\rm blockdiag} \begin{bmatrix} V &V\end{bmatrix} x + \begin{bmatrix} \Pi^T & \!\!\!\!|\!\!\!\! & 0_{1 \times n}  \end{bmatrix}^T \notag
\end{align}

\noindent where $V \in R^{n \times n}$ and $\Pi \in R^n$ have elements given by
\begin{equation*}
V_{ij} = \left\{
\begin{aligned}
-&1 &&i+j = n \\
 &1 && i = n\\
 &0 &&{\rm otherwise}
\end{aligned} \right. \,
\Pi_{i} = \left\{
\begin{aligned}
 &\pi &&i = (n\!-\!1)/2 \\
-&\pi &&i= \{(n\!+\!1)/2, n\} \\
 &0 &&{\rm otherwise}
\end{aligned} \right.
\end{equation*}

\noindent where $0_{i \times j} \in R^{i \times j}$ is the matrix of zeros.\
 \begin{figure}[b!]
 \centering
 \psfrag{A}[][]{\scriptsize {$\encircled{1}$}}
 \psfrag{B}[][]{\scriptsize {$\encircled{2}$}}
 \psfrag{C}[][]{\scriptsize {$\encircled{3}$}}
 \psfrag{D}[][]{\scriptsize {$\encircled{4}$}}
 \psfrag{E}[][]{\scriptsize {initial conditions for next step}}
 \psfrag{F}[][]{\scriptsize {single support phase: swing phase}}
 \psfrag{G}[][]{\scriptsize {double support phase: foot-ground interaction and leg interchange}}
 \includegraphics[width=0.9\linewidth]{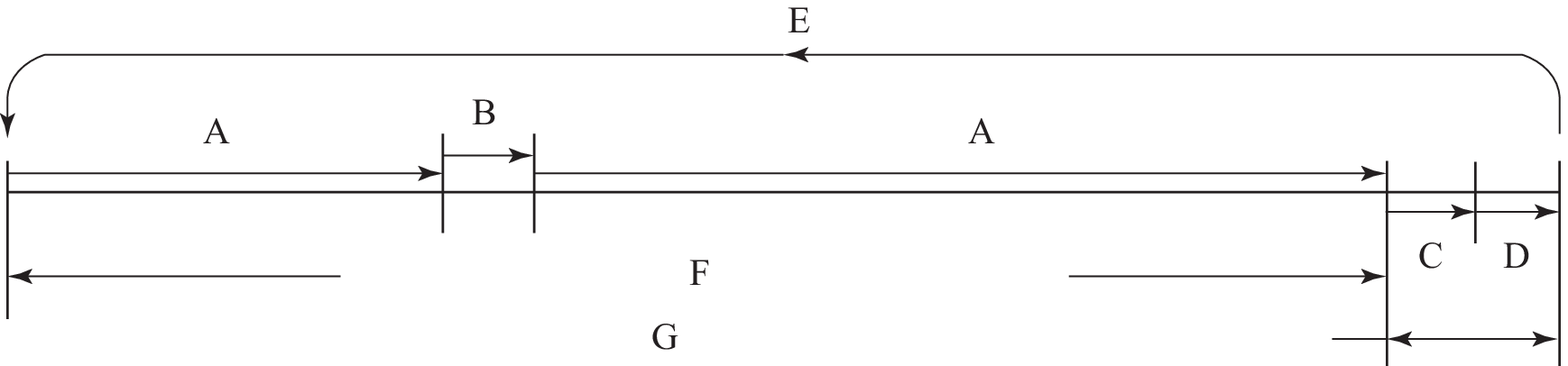}
 \caption{Hybrid dynamics of biped over a step with a single impulsive actuation during the swing phase. The different components are:\newline
 $\protect\encircled{1}$: continuous-time dynamics, \newline
 $\protect\encircled{2}$: jump in states due to impulsive actuation, \newline
 $\protect\encircled{3}$: jump in states due to foot-ground in interaction, and \newline
 $\protect\encircled{4}$: change of states due to coordinate relabelling.}
 \label{Fig2}
 \end{figure}

\subsection{Hybrid Dynamic Model} \label{sec53}

The hybrid dynamics of the gait is described as follows:
\begin{equation} \label{eq13}
\mathcal{D}: \left\{
\begin{aligned}
&\dot x = f(x), & &x \not\in \mathcal{S}, u_{\mathcal{I}} = 0  &\quad&\encircled{1} \\
&x^+ = x^- + \Delta x_{\mathcal{I}}, & &x^- \not\in \mathcal{S}, u_{\mathcal{I}} \neq 0  &\quad&\encircled{2} \\
&x^+ = x^- + \Delta x_{\rm g}, & &x^- \in \mathcal{S}_1  &\quad&\encircled{3} \\
&x^+ = \mathcal{R}(x^-), & &x^- \in \mathcal{S}_2  &\quad&\encircled{4}
\end{aligned} \right.
\end{equation}

\noindent where
\begin{subequations} \label{eq14}
\begin{align}
\mathcal{S}_1 &\triangleq \{x \in \mathcal{Q}^n \times R^n : \gamma_y = 0, \dot \gamma_y < 0\} \label{eq14a} \\
\mathcal{S}_2 &\triangleq \{x \in \mathcal{Q}^n \times R^n : \gamma_y = 0, \dot \gamma = 0\} \label{eq14b}
\end{align}
\end{subequations}

\noindent and $\mathcal{S} \triangleq \mathcal{S}_1 \cup \mathcal{S}_2$ is the set of states during the double-support phase. The different components of the hybrid dynamics $\mathcal{D}$ over a step are illustrated with the help of Fig.\ref{Fig2} for a single impulsive actuation during the swing phase.\

\section{Energy-Conserving Gaits} \label{sec3}

We define an energy-conserving gait as one where the biped has equal mechanical energy at the beginning and end of the swing phase, and there is no loss of energy when the swing leg comes in contact with the ground. The conditions for a gait to be energy conserving are expressed in terms of $\theta_j$ and $\dot\theta_j$, $j = 1, 2, \cdots, n$, to aid physical understanding. Let $(.)^i$ and $(.)^f$ denote the value of $(.)$ at the beginning and end of the swing phase.\

\subsection{Sufficient Conditions for an Energy-Conserving Gait} \label{sec31}

A gait will be energy-conserving if the following conditions are satisfied:
\begin{itemize}
\item The potential energy at the beginning and end of the swing phase are identical, \emph{i.e.}
\begin{equation} \label{eq15}
\theta_j^f = -\theta_j^i, \quad  j = 1, 2, \cdots, n
\end{equation} 

\item The kinetic energy at the beginning and end of the swing phase are identical, \emph{i.e.}
\begin{equation} \label{eq16}
\dot\theta_j^f = \dot\theta_j^i, \quad j=1, 2, \cdots, n 
\end{equation}

\item The interaction between the swing leg and the ground at the end of the swing phase is impact-free, \emph{i.e.}
\begin{equation} \label{eq17}
\dot\gamma^f = -\!\!\! \sum_{\substack{j = 1\\ j \neq (n+1)/2}}^n \!\!\!\ell_j \begin{bmatrix} \cos\theta_j^f \\ \sin\theta_j^f \end{bmatrix} \dot\theta_j^f = 0
\end{equation}

\noindent where link $(n\!+\!1)/2$ is the torso and hence excluded.
\end{itemize}

\subsection{Gait Design Using VHCs} \label{sec32}

To satisfy \eqref{eq15}, we first make the assumption:
\begin{assumption} \label{asm2}
The orientation of the passive link, link $1$, is symmetric with respect to the vertical at the beginning and the end of the swing phase, \emph{i.e.}, $\theta_1^f = -\theta_1^i$.
\end{assumption}

\noindent Next, we impose the following VHCs \cite{maggiore_virtual_2013, kant_orbital_2020} on the actuated joint trajectories during swing phase
\begin{equation} \label{eq18}
\theta_j =  a_j\theta_1 + k_j\pi + f_j^o(\theta_1), \quad j = 2, 3, \cdots, n
\end{equation}
where $a_j \in R$ and $k_j \in Z$ are constants, and $f_j^o(\theta_1)$ is an odd function. The VHCs, along with Assumption \ref{asm2}, ensure that \eqref{eq15} is satisfied for all actuated joint angles. This is true because $k_j\pi = -k_j\pi\,\, \forall k_j \in Z$ and $\theta_j \in S^1$.

Taking the time derivative of \eqref{eq18}, we obtain
\begin{equation} \label{eq19}
\dot\theta_j = \left[a_j + \frac{d f_j^o}{d \theta_1} \right] \dot\theta_1, \quad j = 2, 3, \cdots, n
\end{equation}

\noindent In the above equation, $ \left[a_j + (d f_j^o/d \theta_1) \right]$ is an even function of $\theta_1$; therefore \eqref{eq16} will be satisfied if $\dot\theta_1$ is an even function of $\theta_1$. We will show in section \ref{sec34} that this requirement can be satisfied for the choice of VHCs in \eqref{eq18}. The condition in \eqref{eq17} can be satisfied by choosing the VHC parameters judiciously.\

\subsection{Constraints on VHC Parameters} \label{sec33}

For a single-step periodic gait, we impose the condition that the biped has the same configuration at the beginning of the swing phase. This implies that the biped configuration at the beginning and end of the swing phase is symmetric about the vertical passing through the stance foot, \emph{i.e.},
\begin{subequations} \label{eq20}
\begin{alignat}{3} 
\theta_{n - j + 1}^f &= \theta_j^i - \pi \quad \forall &&j &&= 1, 2, \cdots, (n\!-\!1)/2 \label{eq20a} \\
\theta_j^f &= \theta_j^i &&j &&= (n\!+\!1)/2 \label{eq20b}
\end{alignat}
\end{subequations}

\noindent Using \eqref{eq15}, we get
\begin{subequations} \label{eq21}
\begin{alignat}{3}
\theta_{n - j + 1}^i &= - \theta_j^i + \pi \quad \forall &&j &&= 1, 2, \cdots, (n\!-\!1)/2 \label{eq21a} \\
\theta_j^i &= -\theta_j^i &&j &&= (n\!+\!1)/2 \label{eq21b}
\end{alignat}
\end{subequations}

\noindent From \eqref{eq21b}, it follows
\begin{equation} \label{eq22}
\theta_j^i = 0 \quad j = (n\!+\!1)/2
\end{equation}

\noindent The joint velocities must satisfy
\begin{equation} \label{eq23}
\dot\theta_{n - j + 1}^f = \dot\theta_j^i \quad \forall j = 1, 2, \cdots, (n\!-\!1)/2
\end{equation}

\noindent Using \eqref{eq16}, we get
\begin{equation} \label{eq24}
\dot\theta_{n - j + 1}^i = \dot\theta_j^i \quad \forall j = 1, 2, \cdots, (n\!-\!1)/2
\end{equation}

\noindent By substituting \eqref{eq15} and \eqref{eq16} in \eqref{eq17}, we obtain
\begin{equation} \label{eq25}
\sum_{\substack{j = 1\\ j \neq (n+1)/2}}^n \!\!\!\ell_j \begin{bmatrix} \,\,\,\,\,\cos\theta_j^i \\ -\sin\theta_j^i \end{bmatrix} \dot\theta_j^i = 0
\end{equation}

\noindent Since the legs are kinematically identical, substitution of \eqref{eq21} and \eqref{eq24} into \eqref{eq25} yields
\begin{equation} \label{eq26}
\sum_{j = 1}^{(n-1)/2} \ell_j \sin\theta_j^i \,\dot\theta_j^i = 0
\end{equation}

\noindent It should be noted that the first equation in \eqref{eq25} is trivially satisfied.\

\begin{remark} \label{rem1}
It follows from \eqref{eq19} that the constraints in \eqref{eq24} and \eqref{eq26} are independent of the value of $\dot\theta_1^i$. Thus, the $n+1$ constraints in \eqref{eq21a}, \eqref{eq22}, \eqref{eq24} and \eqref{eq26} depend only on the choice of $\theta_1^i$. 
\end{remark}

\subsection{Zero Dynamics of Energy-Conserving Gait} \label{sec34}
In terms of the generalized coordinates, the VHCs in \eqref{eq18}, with parameters chosen subject to the constraints in \eqref{eq21a}, \eqref{eq22}, \eqref{eq24} and \eqref{eq26}, are expressed as follows:
\begin{equation} \label{eq27}
\rho(q) = q_1 - \Phi(q_2) = 0, \quad \Phi: S^1 \to \mathcal{Q}^{n-1}
\end{equation}

\noindent The corresponding constraint manifold $\mathcal{C}$ is given by:
\begin{equation} \label{eq28}
\mathcal{C} = \left\{ (q, \dot q) : q_1 = \Phi(q_2), \dot q_1 = \left[\frac{\partial \Phi}{\partial q_2}\right] \dot q_2 \right\}
\end{equation}

\begin{remark} \label{rem2}
Since the generalized coordinates are linear combinations of $\theta_j$'s, and linear combinations of odd functions are odd, the function $\Phi(q_2)$ is odd in $q_2$, \emph{i.e}, $\Phi(q_2) = \Phi(-q_2)$, which satisfies \cite[Assumption 2]{kant_orbital_2020}.\
\end{remark}

\begin{remark} \label{rem3}
The VHCs are chosen such that $M_{12}^T (\partial \Phi/ \partial q_2) + M_{22} \neq 0$; this ensures that the VHC in \eqref{eq27} is regular and $\mathcal{C}$ is stabilizable \cite[Remark 2]{kant_orbital_2020}.
\end{remark}

For initial conditions on  $\mathcal{C}$, the continuous control $u_c$ in \cite{kant_orbital_2020} enforces the VHC and renders  $\mathcal{C}$ controlled invariant. On $\mathcal{C}$, the system satisfies $q_1 \equiv \Phi(q_2)$. By substituting \eqref{eq27} in \eqref{eq1b}, we get the swing phase zero dynamics \cite{kant_orbital_2020, byrnes_asymptotic_1991}:
\begin{equation} \label{eq29}
\ddot q_2 = \alpha_1(q_2) + \alpha_2(q_2) \dot q_2^2
\end{equation}

\begin{figure}[b!]
 \centering
 \psfrag{A}[][]{\scriptsize {$\encircled{1}$}}
 \psfrag{B}[][]{\scriptsize {$\encircled{2}$}}
 \psfrag{C}[][]{\scriptsize {$\encircled{3}$}}
 \psfrag{D}[][]{\scriptsize {$\encircled{4}$}}
 \psfrag{E}[][]{\scriptsize {initial conditions for next step}}
 \psfrag{F}[][]{\scriptsize {single support phase: swing phase}}
 \psfrag{G}[][]{\scriptsize {double support phase: leg interchange}}
 \includegraphics[width=0.9\linewidth]{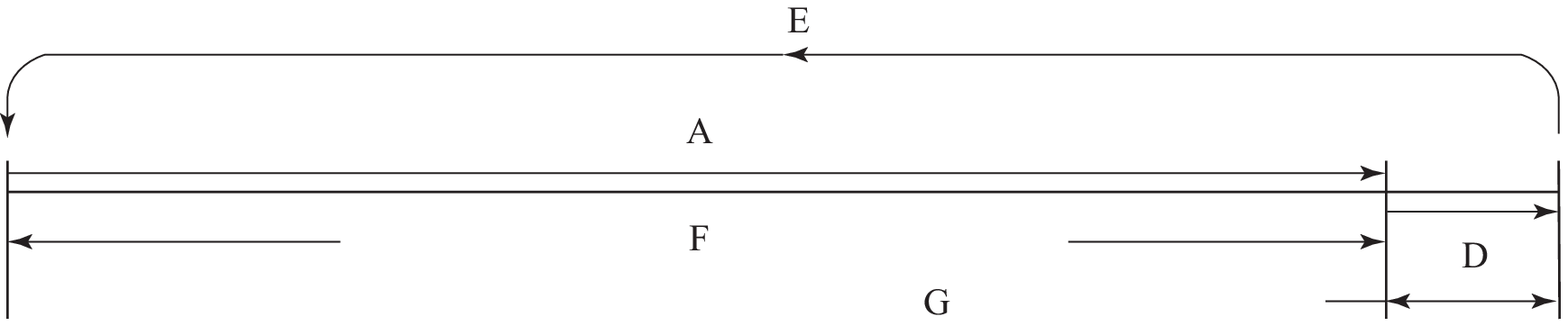}
 \caption{Hybrid dynamics of biped over a step for an energy-conserving gait; it is a simpler version of the dynamics shown in Fig.\ref{Fig2}. The different components are:\newline
 $\protect\encircled{1}$: continuous-time dynamics, \newline
 $\protect\encircled{4}$: change of states due to coordinate relabelling.}
 \label{Fig3}
 \end{figure}

\noindent It can be shown that $\ddot q_2$ is an odd function of $q_2$ since the VHCs in \eqref{eq27} are odd. This implies that both $\alpha_1$ and $\alpha_2$ are odd functions of $q_2$; and $\dot q_2$ is either an even function or an odd function of $q_2$, but not neither. The zero dynamics in \eqref{eq29}
has an integral of motion $E(q_2, \dot q_2)$ \cite{shiriaev_constructive_2005, maggiore_virtual_2013}, and its qualitative properties can be described by a potential energy function $\mathcal{P}(q_2)$, which has minimum and maximum values $\mathcal{P}_{\rm min}$ and $\mathcal{P}_{\rm max}$ \cite{maggiore_virtual_2013}.  For an energy level set $E(q_2, \dot q_2) = c$, $c \in (\mathcal{P}_{\rm min}, \mathcal{P}_{\rm max})$ corresponds to a periodic orbit where $\dot q_2$ changes sign periodically ($\dot q_2$ is an odd function of $q_2$) and $c > \mathcal{P}_{\rm max}$ corresponds to an orbit where $\dot q_2$ does not change sign ($\dot q_2$ is an even function of $q_2$) \cite{maggiore_virtual_2013, kant_orbital_2020}. 
If $\dot q_2$ changes sign when $q_2$ changes sign, the biped will not be able to complete a step. Therefore, a feasible biped gait is a periodic orbit where $\dot q_2$ does not change sign. This will be accomplished through proper choice of initial conditions that result in $c > \mathcal{P}_{\rm max}$.\

A system trajectory in $\mathcal{C}$ satisfies $\dot\gamma^f = 0$ $\forall x \in \mathcal{C}$. Thus, $\mathcal{C} \cap \mathcal{S}_1 = \emptyset$ and consequently $\mathcal{C} \cap \mathcal{S} = \mathcal{C} \cap \mathcal{S}_2$. A trajectory evolving in $\mathcal{C}$ in the single-support phase will intersect $\mathcal{S}_2$ in the double-support phase. This results in a discontinuous jump in states described by \eqref{eq12}. Although the trajectory may leave $\mathcal{C}$ during the jump, it can be shown that the new states lie in $\mathcal{C}$, \emph{i.e.}, $\mathcal{C}$ is invariant under relabelling of the states:
\begin{equation} \label{eq30}
\mathcal{R}(\mathcal{C} \cap \mathcal{S}_2) \subset \mathcal{C}
\end{equation}

\noindent The components of the hybrid dynamics over a step for an energy-conserving gait are shown in Fig.\ref{Fig3}; the evolution of the system trajectory is shown in Fig.\ref{Fig4}.\

\section{Stabilization of an Energy-Conserving Gait}

\subsection{Orbit Describing an Energy-Conserving Gait}

An energy-conserving gait, defined by the VHCs in \eqref{eq27}, is described by the hybrid orbit:
\begin{equation} \label{eq31}
\mathcal{O}^* = \mathcal{O}^*_\mathcal{C} \cup \mathcal{O}^*_\mathcal{R}
\end{equation}

 \begin{figure}[t!]
  \vspace{0.10in}
 \centering
 \psfrag{A}[][]{\scriptsize {$\encircled{4}$}}
 \psfrag{B}[][]{\small {$\mathcal{S}_2$}}
 \psfrag{C}[][]{\small {$\mathcal{C}$}}
 \psfrag{D}[][]{\scriptsize {$x^i\!=\!x^+$}}
 \psfrag{E}[][]{\scriptsize {$x^f\!=\!x^-$}}
 \psfrag{F}[][]{\scriptsize {$\encircled{1}$}}
 \includegraphics[width=0.5\linewidth]{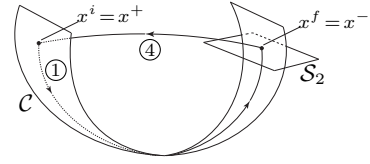}
 \caption{Evolution of system trajectory during an energy-conserving gait.}
 \label{Fig4}
 \vspace{-0.20in}
 \end{figure}
 
\noindent where
\begin{subequations} \label{eq32}
\begin{align} 
\mathcal{O}^*_\mathcal{C} &= \left\{ x \in \mathcal{C} : E(q_2, \dot q_2) = c^* \right\} \qquad c^* > \mathcal{P}_{\rm max} \label{eq32a} \\
\mathcal{O}^*_\mathcal{R} &= \left\{ x^-, x^+ : x^- \in \mathcal{O}^*_\mathcal{C} \cap \mathcal{S}_2, x^+ = \mathcal{R}(x^-) \in \mathcal{O}^*_\mathcal{C}\right\} \label{eq32b}
\end{align}
\end{subequations}

%

\noindent The orbit $\mathcal{O}^*$ is stabilized using the ICPM approach \cite{kant_orbital_2020}, whose efficacy has been demonstrated for both continuous and hybrid orbits \cite{kant_orbital_2020, kant_juggling_2022}.\

\subsection{Poincar\'e Map}

To stabilize the orbit $\mathcal{O}^*$ from any point in its neighborhood, we describe the hybrid dynamics in \eqref{eq13} using a discrete-time map. To this end, we define the Poincar\'e section:
\begin{equation} \label{eq33}
\Sigma = \{x \in \mathcal{Q}^n \times R^n : q_2 = q_2^*, \dot q_2 < 0\}
\end{equation}

\noindent The states on $\Sigma$ are:
\begin{equation} \label{eq34}
z = \begin{bmatrix} q_1^T & \dot q^T \end{bmatrix}^T, \quad z \in \mathcal{Q}^{n-1} \times R^n
\end{equation}

\noindent We assume that impulsive actuation is applied when the system trajectory intersects $\Sigma$. If $z(k)$ denotes the states immediately prior to application of $\mathcal{I}$, the hybrid dynamics of the impulse-controlled system can be expressed as
\begin{equation} \label{eq35}
z(k+1) = \mathbb{P} [z(k), \mathcal{I}(k)]
\end{equation}

\noindent The map $\mathbb{P}$ captures the dynamics between subsequent intersections of the system trajectory with $\Sigma$. It is comprised of the components $\encircled{2}$,
$\encircled{1}$, $\encircled{3}$, $\encircled{4}$, and $\encircled{1}$, described in \eqref{eq13} and depicted in Fig.\ref{Fig2}.


\subsection{Orbital Stabilization}
If $x \in \mathcal{O}^*$, the system trajectory is restricted to $\mathcal{O}^*$ under continuous control $u_c$. The intersection of $\mathcal{O}^*$ with $\Sigma$ is therefore a fixed point $z(k) = z^*$,
$\mathcal{I}(k) = 0$ of $\mathbb{P}$
\begin{equation} \label{eq36}
z^* = \mathbb{P} (z^*, 0)
\end{equation}

\noindent If $x \not\in {O}^*$, $u_c$ does not guarantee convergence of the trajectory to $\mathcal{O}^*$, and the impulsive inputs $\mathcal{I}(k)$ are used to asymptotically stabilize the fixed point $z^*$, and consequently the orbit $\mathcal{O}^*$ \cite{khalil_nonlinear_2002, kant_orbital_2020}. To this end, we linearize the map $\mathbb{P}$ about $z(k) = z^*$ and $\mathcal{I}(k) = 0$ as follows:
\begin{equation} \label{eq37}
e(k+1) = \mathcal{A} e(k) + \mathcal{B} \mathcal{I}(k),\quad  e(k) \triangleq z(k) - z^* 
\end{equation}

\noindent where
\begin{equation} \label{eq38}
\begin{split}
\mathcal{A} &\triangleq  \left[\nabla_{z}\mathbb{P}(z, \mathcal{I})\right]_{z = z^*\!, \,\mathcal{I} = 0} \\
\mathcal{B} &\triangleq  \left[\nabla_{\mathcal{I}}\mathbb{P}(z, \mathcal{I}) \right]_{z = z^*\!, \,\mathcal{I} = 0}
\end{split}
\end{equation}

\noindent The matrices $\mathcal{A} \in R^{(2n-1)\times(2n-1)}$ and $\mathcal{B} \in R^{(2n-1)\times(n-1)}$ can be computed numerically. If $(\mathcal{A}, \mathcal{B})$ is controllable, the orbit $\mathcal{O}^*$ can be stabilized by the discrete feedback:
\begin{equation}\label{eq39}
\mathcal{I}(k) = \mathcal{K}e(k)
\end{equation}

\noindent where $\mathcal{K}$ is chosen such that the eigenvalues of $(\mathcal{A}+\mathcal{B}\mathcal{K})$ lie inside the unit circle.\

\begin{figure}[t!]
\vspace{0.10in}
\centering
\psfrag{A}[][]{\scriptsize {$\encircled{1}$}}
\psfrag{B}[][]{\scriptsize {$\encircled{4}$}}
\psfrag{C}[][]{\scriptsize {$\encircled{2}$}}
\psfrag{D}[][]{\scriptsize {$\encircled{3}$}}
\psfrag{E}[][]{\small {$\Sigma$}}
\psfrag{F}[][]{\small {$\mathcal{O}^*$}}
\psfrag{G}[][]{\small {$z^*$}}
\includegraphics[width=0.58\linewidth]{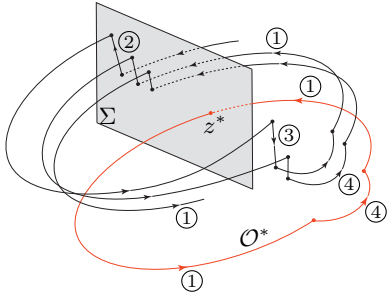}
\caption{Schematic of the ICPM approach to orbital stabilization of an energy-conserving gait. The desired orbit is shown in red. The different components of the hybrid dynamics, namely,
$\protect\encircled{1}$, $\protect\encircled{2}$, $\protect\encircled{3}$ and $\protect\encircled{4}$ are described by \eqref{eq13}.}
\label{Fig5}
\vspace{-0.20in}
\end{figure}

The stabilization of $\mathcal{O}^*$ using the ICPM approach \cite{kant_orbital_2020, kant_juggling_2022} is explained with the help of Fig.\ref{Fig5}. The desired orbit $\mathcal{O}^*$ (shown in red), intersects
$\Sigma$ at the fixed point $z^*$; it corresponds to an energy-conserving gait where the states undergo a single discontinuous jump due to coordinate relabelling - see Fig.\ref{Fig4}. For a trajectory not on $\mathcal{O}^*$ (shown in black), there is a discontinuous jump in states on $\Sigma$ due to application of $\mathcal{I}(k)$. A second discontinuous jump occurs at the time of foot-ground interaction; this is immediately followed by the discontinuous jump due to coordinate relabelling. The input $\mathcal{I}(k)$ in \eqref{eq39} guarantees asymptotic convergence of a system trajectory to $\mathcal{O}^*$.\

\section{Gait Stabilization for Five-Link Biped} \label{sec5}

\subsection{Gait Selection} \label{sec52}

For $n=5$, \eqref{eq26} has the form
\begin{equation} \label{eq40}
\ell_1 \sin\theta_1^i \,\dot\theta_1^i + \ell_2 \sin\theta_2^i \,\dot\theta_2^i = 0
\end{equation}

\noindent which permits us to obtain a non-trivial gait. We seek a gait where the odd functions $f_j^o(\theta_1)$ in \eqref{eq18} are sinusoidal. Thus \eqref{eq18} can be rewritten as:
\begin{equation} \label{eq41}
\theta_j =  a_j\theta_1 + k_j\pi + \mathcal{G}_j \sin(\mathcal{H}_j \theta_1), \quad j = 2, 3, 4, 5 
\end{equation}

\noindent where $\mathcal{G}_j, \mathcal{H}_j \in R$, $j = 2, 3, 4, 5$, are constants. For a certain choice of $\theta_1^i$, the 16 VHC parameters in \eqref{eq41} should satisfy the six constraints in \eqref{eq21a}, \eqref{eq22}, \eqref{eq24} and \eqref{eq40}. For the kinematic and dynamic parameters in Table \ref{tab1}, and the choice $\theta_1^i = \pi/8$, a set of feasible VHC parameters which ensure a realistic walking motion are listed in Table \ref{tab2}.

Using \eqref{eq41} and the parameters in Table \ref{tab2}, \eqref{eq27} can be expressed as:
\begin{equation} \label{eq42}
\begin{split}
\rho(q) &= q_1 - \Phi(q_2) = 0, \quad \Phi: S^1 \to \mathcal{Q}^4 \\
\Phi(q_2) \!&=\! 
\begin{bmatrix} 
-0.4500 q_2 + 0.2717 \sin(8 q_2) \\
-0.5500 q_2 - 0.6717 \sin(8 q_2) \\
-0.5500 q_2 + \pi + 0.5342 \sin(8 q_2) \\
-1.1333 q_2 -0.1342 \sin(8 q_2) -0.3795 \sin(10 q_2)
\end{bmatrix}
\end{split}
\end{equation}

\noindent The above VHC satisfies the conditions in remarks \ref{rem2} and \ref{rem3}, and is enforced by the continuous control $u_c$ in \cite{kant_orbital_2020}\footnote{The expression for $u_c$ uses the gains $k_p = 750 I_{4}$ and $k_d = 25 I_{4}$, where $I_{n} \in R^{n \times n}$ is the identity matrix; these gains were chosen to ensure rapid convergence of the trajectories to the constraint manifold $\mathcal{C}$.}. The desired orbit $\mathcal{O}^*$, defined by \eqref{eq31} and \eqref{eq32} is chosen to be the one which passes through $(q_2, \dot q_2) = (\pi/8, -5\pi/3)$, for which the sign of $\dot q_2$ does not change.\

\begin{remark} \label{rem4}
There is significant flexibility in the choice of VHCs and their parameters.\
\end{remark}

\begin{table}[t!]
\caption{Kinematic and dynamic parameters of five-link biped}
\centering
\begin{tabular}{|l|c|c|c|c|} \hline
\multicolumn{1}{|c|}{$j$} &$\ell_j$ (m) & $d_j$ (m) & $m_j$ (kg) &$J_j$ (kg m$^2$) \\ \hline
1, 5 &0.5000 &0.2500 &0.4000 &0.0083 \\ \hline
2, 4 &0.5500 &0.2750 &0.4500 &0.0113 \\ \hline
3 (torso) &0.6000 &0.3000 &0.5500 &0.0165 \\ \hline
\end{tabular}
\label{tab1}
\end{table}
\begin{table}[t!]
\caption{VHC parameters for energy-conserving gait}
\centering
\begin{tabular}{|l|r|c|r|c|} \hline
\multicolumn{1}{|c|}{$j$} &\multicolumn{1}{c|}{$a_j$} & $k_j$ & \multicolumn{1}{c|}{$\mathcal{G}_j$} &$\mathcal{H}_j$ \\ \hline
2 &$0.5500$ &0 &$0.2717$ &$8$ \\ \hline
3 (torso) &0.0000 &0 &$-0.4000$ &$8$ \\ \hline
4 &$-0.5500$ &1 &$0.1342$ &$8$ \\ \hline
5 &$-1.6833$ &1 &$-0.3795$ &$10$ \\ \hline
\end{tabular}
\label{tab2}
\vspace{-0.15in}
\end{table}

\subsection{Stabilization of $\mathcal{O}^*$} \label{sec52}

We choose the following Poincar\'e section:
\begin{equation} \label{eq43}
\Sigma = \{x \in \mathcal{Q}^5 \times R^5 : q_2 = \pi/16, \dot q_2 < 0\}
\end{equation}

\noindent on which the states are denoted by $z, z \in \mathcal{Q}^4 \times R^5$, defined in \eqref{eq34}. The intersection of $\mathcal{O}^*$ with $\Sigma$ in \eqref{eq43} is the fixed point $z^*$ of the map $\mathbb{P}$, and satisfies \eqref{eq36}. The matrices $\mathcal{A} \in R^{9 \times 9}$ and $\mathcal{B} \in R^{9 \times 4}$ in \eqref{eq38} are computed numerically \cite{kant_orbital_2020}; their expressions are not provided here for brevity. It is seen that the eigenvalues of $\mathcal{A}$ do not all lie within the unit circle, but the pair $(\mathcal{A}, \mathcal{B})$ is controllable. The gain matrix $\mathcal{K}$ in \eqref{eq39}, which asymptotically stabilizes $\mathcal{O}^*$, is obtained using LQR; the gain matrices were chosen as\
\begin{equation*}
Q = {\rm blockdiag} \begin{bmatrix} I_{4} &1.5 I_{5} \end{bmatrix},\quad R = I_{4}
\end{equation*}

\subsection{Simulation results} \label{sec53}
\begin{figure}[t!]
\vspace{0.10in}
\centering
\psfrag{A}[][]{\footnotesize {$\rho_1$}}
\psfrag{B}[][]{\footnotesize {$\rho_2$}}
\psfrag{C}[][]{\footnotesize {$\rho_3$}}
\psfrag{D}[][]{\footnotesize {$\rho_4$}}
\psfrag{E}[][]{\footnotesize {time (s)}}
\psfrag{F}[][]{\footnotesize {$q_2$ (rad)}}
\psfrag{I}[][]{\footnotesize {$t \leq 3$ s}}
\psfrag{J}[][]{\footnotesize {$t > 3$ s}}
\psfrag{K}[][]{\footnotesize {$\mathcal{O}^*$}}
\psfrag{G}[][]{\footnotesize {$\dot q_2$ (rad/s)}}
\psfrag{L}[][]{\footnotesize {$\|e(k)\|_2$}}
\psfrag{M}[][]{\footnotesize {$k$}}
\includegraphics[width=0.95\linewidth]{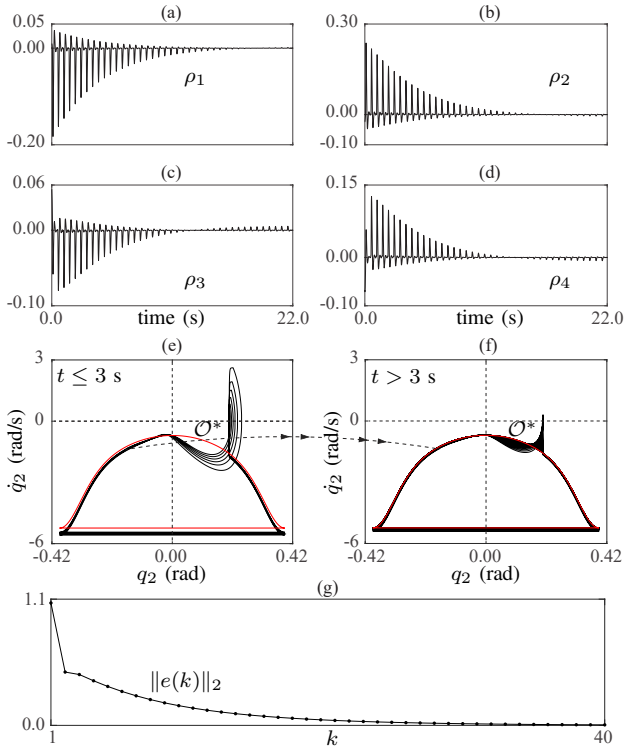}
\caption{Orbital stabilization of an energy-conserving gait using the ICPM approach: (a)-(d) show the components of $\rho(q)$, (e) and (f) show the phase portrait of the passive coordinate, and (g) shows the norm of the error in states on the Poincar\'e section.}
\label{Fig6}
 \vspace{-0.20in}
\end{figure}

For an arbitrary set of initial conditions not lying on $\mathcal{O}^*$, simulation results of the ICPM approach to gait stabilization are shown in Fig.\ref{Fig6} for 40 steps, that corresponds to a duration of approx. 22 s. For the biped walking with the desired energy-conserving gait, a single step is completed in 0.5646 s under the continuous control $u_c$. The impulsive control in \eqref{eq39} is realized in the simulations using high-gain feedback $u_{\rm hg}$, as in \cite{kant_orbital_2020}\footnote{The expression for $u_{\rm hg}$ uses $\Lambda = I_{4}$ and $\mu = 0.0005$; the high-gain feedback is terminated when the norm of the error in the active joint velocities reduces below $0.0001$.}; the components of $\rho(q)$ are plotted in Fig.\ref{Fig6}(a)-(d); these plots demonstrate convergence of system trajectories to the constraint manifold $\mathcal{C}$. The phase portrait of the passive coordinate $q_2$ is shown in Fig.\ref{Fig6}(e) for $t \leq 3$ s and Fig.\ref{Fig6}(f) for $t > 3$ s. The desired orbit $\mathcal{O}^*$ is shown in red in both Fig.\ref{Fig6}(e) and Fig.\ref{Fig6}(f); it is observed that system trajectories asymptotically converge to $\mathcal{O}^*$. Finally, $\|e(k)\|_2$, $k = 1, 2, \cdots, 40$, is plotted in Fig.\ref{Fig6}(g), which demonstrates asymptotic convergence of the states $z(k)$ on $\Sigma$ to $z^*$.\

\section{Conclusion} \label{sec6}

For point-foot planar bipeds, a method for designing and stabilizing impact-free, energy-conserving gaits is presented. VHCs are used to design energy-conserving gaits, which are hybrid orbits with a single discontinuity due to coordinate relabelling. A system trajectory that does not lie on such an orbit is subject to discontinuities due to impulsive disturbances arising from foot-ground interaction. To stabilize a desired orbit, we use the ICPM approach, wherein impulsive inputs are applied when the system trajectory intersects a chosen Poincar\'e section. As the system trajectory converges to the desired orbit, the magnitudes of both the impulsive inputs and impulsive disturbance converge to zero. This approach differs from previous approaches in that impulsive inputs are included in the set of admissible controls. The efficacy of the ICPM approach is demonstrated through a case study of a five-link biped. Future work will focus on search for a wider class of VHCs, and optimal selection of the Poincar\'e section on which to apply impulsive inputs.\
%
%
\bibliographystyle{IEEEtran}      
\bibliography{ref}   

\end{document}